\documentclass[journal,12pt,onecolumn,draftclsnofoot]{IEEEtran}
\usepackage{tikz}
\usetikzlibrary{positioning}
\usepackage{subcaption}

\newtheorem{thm}{Theorem}

\newtheorem{defn}{Definition}
\newcommand{\ds}{\displaystyle}
\newcommand{\la}{\lambda}
\newcommand{\SSS}{\mathcal{S}}
\newcommand{\LL}{\mathcal{L}}
\newcommand{\A}{\mathcal{A}}

\ifCLASSINFOpdf
\else
\fi
\usepackage{amsmath}
\begin{document}
%
\title{3-Dimensional Optical Orthogonal Codes with Ideal Autocorrelation-Bounds and Optimal Constructions}
%
%
%

\author{Tim~Alderson,~\IEEEmembership{}
\thanks{T. Alderson is with the Department
of Mathematics and Statistics, University of New Brunswick Saint John, Saint John,
NB, E2L 4L5 Canada e-mail: Tim@unb.ca.}
}

%
%

\markboth{Manuscript 2016}%
{ : Bare Demo of IEEEtran.cls for IEEE Journals}
%



\maketitle

\begin{abstract}
Several new constructions of 3-dimensional optical orthogonal codes are presented here. In each case the codes have ideal autocorrelation $\mathbf{  \la_a=0} $, and in all but one case a cross correlation of $ \mathbf{\la_c=1} $. All codes produced are optimal with respect to the applicable Johnson bound either presented or developed here. Thus, on one hand the codes are as large as possible, and on the other, the bound(s) are shown to be tight.  All codes are constructed by using a particular automorphism (a Singer cycle) of $ \mathbf{ PG(k,q)} $, the finite projective geometry of dimension $ k $ over the field of order $ \mathbf{q} $, or by using an affine analogue in $ AG(k,q) $.   

\end{abstract}

\begin{IEEEkeywords}
3-D code, 3-D OOC, Optical Orthogonal Codes, Johnson bound, finite projective geometries, PG(k,q), Singer cycle. optimal codes
\end{IEEEkeywords}

%
\IEEEpeerreviewmaketitle

\section{Introduction}
%
%
%
%
\IEEEPARstart{O}{ptical} code division multiple access (OCDMA) continues to be of great interest among multiple access systems due to ease of implementation,  support for asynchronous and secure communication, soft traffic handling capability, and strong performance with high numbers of users \cite{Arnon:2012:AOW:2407921}.  
The work of Salehi \textit{et. al.} \cite{Salehi1989} \cite{Salehi1989a}, spearheaded the use of optical orthogonal codes for OCDMA, and these codes continue to be highly effective over a quarter of a century later.  \\
An $(n,w,\lambda_a,\lambda_c)$-optical orthogonal code (OOC) is a family of (1-dimensional) binary sequences
(codewords) of length $n$, and constant Hamming weight $w$ satisfying the following two conditions:
\begin{itemize}
	\item (auto-correlation property) for any codeword
	$c=(c_0,c_1,\ldots,c_{n-1})$ and for any integer $1\leq t \leq n-1$, we have $\ds
	\sum_{i=1}^{n-1}c_ic_{i+t}\leq \lambda_a$,
	\item (cross-correlation property) for any two distinct codewords
	$c,c'$ and for any integer $0\leq t \leq n-1$, we have $\ds \sum_{i=0}^{n-1}c_ic'_{i+t}\leq
	\lambda_c$,
\end{itemize}
where each subscript is reduced modulo $n$.

An $(n,w,\lambda_a,\lambda_c)$-OOC $ C $ with $\lambda_a=\lambda_c$ is denoted an $(n,w,\lambda)$-OOC.
The number of codewords is the \emph{size} or \textit{capacity} of the code, denoted $ |C| $.  For fixed values of $n$, $w$, $\lambda_a$
and $\lambda_c$, the largest size of an $(n,w,\lambda_a,\lambda_c)$-OOC is denoted
$\Phi(n,w,\la_a,\la_c)$. An $(n,w,\la_a,\la_c)$-OOC is said to be \textit{optimal} if $ |C|= $ $\Phi(n,w,\la_a,\la_c)$
. Optimal OOCs facilitate the largest possible number of asynchronous users to transmit information efficiently and reliably. 

A limitation of 1-D OOCs is that the autocorrelation cannot be zero,  and to maintain minimal autocorrelation of $ 1 $ the code length must increase quite rapidly with the number of users. The 1-D-OOCs spread the input data bits only in the time domain. Technologies such as wavelength-division-multiplexing (WDM) and dense-WDM enable the spreading of codewords  in both space and time \cite{Park1992}, or in wave-length and time \cite{Mendez1995a}.  Hence, codewords may be considered as $\Lambda \times T$ $(0,1)$-matrices. These codes are referred to in the literature as multiwavelength, multiple-wavelength, wavelength-time hopping, and 2-dimensional OOCs (2D-OOCs).
The addition of another dimension allows codes to be constructed with at most a single pulse per row, yielding autocorrelation zero and  thereby improving the OCDMA performance in comparison with 1-D OCDMA. For optimal constructions of 2-D OOC's see \cite{Alderson20111187,1315909,1523307}.  Later, a third dimension was added  which gave an increase the code size and the performance of the code \cite{doi:10.1117/12.238940}. In 3-D OCDMA the optical pulses are spread in three domains space, wave-length, and time, with codes referred to as \textit{space/wavelength/time spreading} codes, or \textit{3-D OOC}.


\subsection{3-D OOCs and Bounds}


We denote by $( \Lambda  \times S\times T,w, \lambda_a, \lambda_c)$  a 3D-OOC with constant weight $w$, $\Lambda$ wavelengths, space spreading length $ S $,
and time-spreading length $T$ (hence, each codeword may be considered as  an $\Lambda  \times S \times T$ binary array).  The autocorrelation and cross correlation of
an $(\Lambda  \times S \times T,w, \lambda_a, \lambda_c)$-3D-OOC have the following properties.

\begin{itemize}
	\item (auto-correlation property) for any codeword
	$A=(a_{i,j,k})$ and for any integer $1\leq t \leq T-1$, we have $\ds
	\sum_{i=0}^{S-1}\sum_{j=0}^{\Lambda-1}\sum_{k=1}^{T-1}a_{i,j,k}a_{i,j, k+t}\leq \lambda_a$,
	\item (cross-correlation property) for any two distinct codewords
	$A=(a_{i,j, k})$, $B=(b_{i,j,k})$ and for any integer $0\leq t \leq T-1$, we have $\ds \sum_{i=0}^{S-1}\sum_{j=0}^{\Lambda-1}\sum_{k=0}^{T-1}a_{i,j,k}b_{i,j, k+t}\leq \lambda_c$,
\end{itemize}
where each subscript is reduced modulo $T$.
There are practical considerations to be made with regard to the implementation of these codes.  First, in optical code-division multiple-access (OCDMA) applications, minimal correlation values  are most desirable. Implementation is simplified (and more cost effective)  when $ \la_a=0 $ \cite{Jugl644393}.  Codes satisfying $ \la_a=0 $ will be said to be \textit{ideal} here. Ideal codes with minimal autocorrelation  $\la_c=1 $ are our main focus. 

A wavelength/time plane is called a \textit{spatial plane}, a space/time plane is called a \textit{wavelength plane}, and a space/wavelength plane is called a \textit{temporal plane}. One way to achieve $ \la_a=0 $ is to select codes with at most one pulse per spatial plane. Such codes are referred to as \textit{at most one pulse per plane} (AMOPP) codes. AMOPP codes of maximal weight $ S $ have a single pulse per spatial plane, and  are referred to as SPP codes. Codes with at most one pulse per  wavelength plane also enjoy zero autocorrelation, and are denoted AMOPW codes.  AMOPW codes of maximal weight $ \Lambda $  are single pulse per wavelength (SPW) codes. Codes with at most one (resp. exactly one)  pulse per temporal plane do not necessarily have $ \la_a=0 $ are referred to as AMOPT and SPT codes respectively.\\
As it is of interest to construct codes with as large cardinality as possible, we now discuss some upper bounds on the size of codes.  
%
%

In order to develop new bounds for codes with ideal autocorrelation we introduce the notion of Hamming correlation. Given two $ 1 $-dimensional codewords over any alphabet, the \textit{Hamming correlation} is the number of non-zero agreements between the two codewords. By an $ (n,w,\lambda)_{m+1} $-code, we denote a code of  length $ n $, with constant weight $ w $,  and maximum Hamming correlation $ \lambda $ over an alphabet of size $ m+1 $   (containing  zero). For binary codes ($ m=1 $) the subscript $ 2 $ is typically dropped. Let $ A(n,w,\lambda)_{m+1}$ denote the maximum size of  an $ (n,w,\lambda)_{m+1} $-code.  The bound of Johnson \cite{MR0137615} establishes the following bound in the binary case. 
\begin{thm}[Johnson Bound \cite{MR0137615}]. \label{thm:JB2}\\
\[ A(n,w,\lambda) \le \left\lfloor
	\frac{  n}{w} \left\lfloor \frac{ (n-1)}{w-1}
	\left\lfloor \cdots \left\lfloor %
	\frac{  (n-\lambda)}{w-\la}\right\rfloor \right\rfloor \cdots \right\rfloor\right.
	.\]
If $ w^2-n\lambda>0 $ then
\[
A(n,w,\lambda)\le \left\lfloor \frac{n(w-\lambda)}{w^2-n\lambda} \right \rfloor.
\]

\end{thm}

Continuing with the binary case, Agrell \textit{et. al.} \cite{Agrell2000} establish the following bound.

\begin{thm}[ \cite{Agrell2000}]. \label{thm:JB2b}\\
	\[ A(n,w,\lambda) \le n 
	 \text{ if } 0  < w^2-n\la \le w-\la \]	
\end{thm}

By identifying alphabet elements with mutually distinct binary strings of length $ m $ and weight at most one, an $ (n,w,\la)_{m+1} $ code can be considered an $ (nm, w, \la) $-code. As such the bounds on binary codes  can easily be adapted to the non-binary case. Moreover, observe that an $ (n,w,\la)_{m+1} $ code attaining the bound $ A(n,w,\la)_{m+1} $ must have a coordinate in which at least $ \frac{w\cdot A(n,w,\la)_{m+1} }{mn} $ codewords have a common nonzero entry. As observed in  \cite{Svanstroem1999}, shortening the code with respect to this coordinate gives a code with at most $ A(n-1,w-1,\la-1)_{m+1} $ codewords.
\begin{thm}[\cite{Svanstroem1999}] \label{thm: Svanstroem}
	\[
A(n,w,\la)_{m+1}\le \left\lfloor \frac{mn}{w} A(n-1,w-1,\la-1)_{m+1}\right \rfloor
	\]
\end{thm}  
Observing that 
$ A(n,w,0)_{m+1} =m\left\lfloor \frac{n-\la}{w-\la} \right\rfloor  $, Theorems \ref{thm:JB2}, \ref{thm:JB2b}, and \ref{thm: Svanstroem} then give the following.


\begin{thm}[Johnson Bound Non-binary]. \label{thm:JB}\\	
	\[   A(n,w,\lambda)_{m+1} \le \left\lfloor
	\frac{m n}{w} \left\lfloor \frac{m(n-1)}{w-1}\left\lfloor 
	\cdots \left\lfloor %
	\frac{ m(n-\lambda)}{w-\la}\right\rfloor \right\rfloor \cdots \right\rfloor\right.
	.\]
If $ w^2>mn\lambda $ then
\[
A(n,w,\lambda)_{m+1}\le \text{ min } \left\{mn,\left\lfloor \frac{mn(w-\lambda)}{w^2-mn\lambda} \right \rfloor \right\}.
\]
	
\end{thm}

We note that the first bound in Theorem \ref{thm:JB} may also be found in \cite{OmraniEK04} with a proof (quite  different from that given here) in \cite{Omrani2012}. 

%

Observe that by choosing a fixed linear ordering, each codeword from an $ (\Lambda\times S\times T,w,\la)$ 3D-OOC $ C $  can be viewed as a binary constant weight ($ w $) code of length $ \Lambda S T $. Moreover, by including the $ T $ distinct cyclic shifts of each codeword we obtain a corresponding constant weight binary code of size $ T \cdot |C| $.  It follows that
\begin{equation}\label{eqn: JB1}
|C|\le \left \lfloor \frac{A(\Lambda S T, w, \la)}{T}\right \rfloor
\end{equation} 

From  the equation (\ref{eqn: JB1}) and Theorem \ref{thm:JB} we obtain the following bounds for 3-D OOCs. 

\begin{thm}[Johnson Bound for 3D OOCs]\label{thm: 3djb2} 
	Let $ C $ be a $ (\Lambda \times S\times T, w, \la) $-OOC. Then
	\begin{align} \label{eqn:3djba} 
	\Phi(C) & \le  \left\lfloor
	\frac{\Lambda S}{w} \left\lfloor \frac{\Lambda S T-1}{w-1}\left\lfloor 
	\cdots \left\lfloor
	\frac{ \Lambda S T -\lambda}{w-\la}\right\rfloor \right\rfloor \cdots \right\rfloor\right.
	. 
	\end{align}
If $ w^2>\Lambda S T \lambda $ then
\begin{equation} \label{eqn:3djbb}
\Phi(C) \le \text{ min } \left\{\Lambda S,\left\lfloor \frac{\Lambda S(w-\lambda)}{w^2- \Lambda S T\lambda} \right \rfloor \right\}.
\end{equation}	
\end{thm}	

We note that the first bound (\ref{eqn:3djba}) may also found in  \cite{Ortiz-Ubarri2011}. 

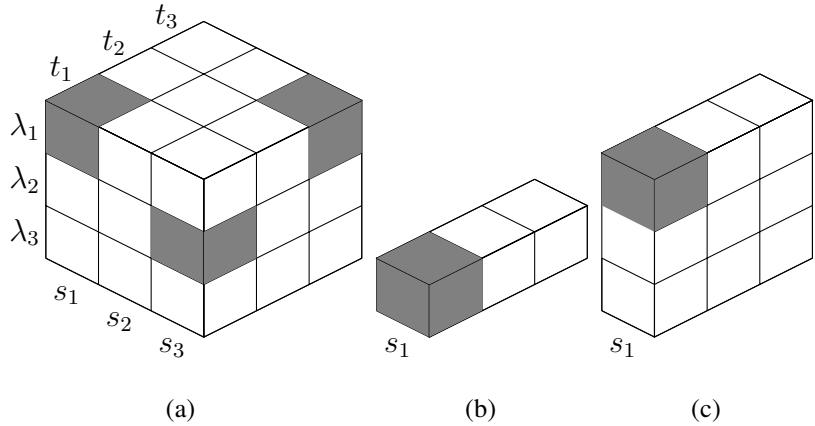
\begin{figure}[htb]
	\centering
	\subcaptionbox{\label{Fig:a}}
{\begin{tikzpicture}[on grid, scale=.7]
\draw[yslant=-0.5,]  (0,0) rectangle +(3,3);
\draw[yslant=-0.5] (0,0) grid (3,3);

\foreach \x in {1,2,3}{
	\node  at (\x -0.6, -\x/2 -0.2) {$ s_{\x} $};
	\node at (-0.4,3-\x +0.5 ) {$ \lambda_{\x} $};
	\node  at (\x -0.7,  \x/2 +3.1) {$ t_{\x} $};		
}

\foreach \x in {2}{
	\foreach \y in {1}{
		\fill[yslant=-0.5, gray] (\x,\y) rectangle +(1,1);
}}
\foreach \x in {0}{
	\foreach \y in {2}{
		\fill[yslant=-0.5, gray] (\x,\y) rectangle +(1,1);
}}
\draw[yslant=0.5] (3,-3) rectangle +(3,3);
\draw[yslant=0.5] (3,-3) grid (6,0);
\foreach \x in {5}{
	\foreach \y in {-1}{
		\fill[yslant=0.5, gray] (\x,\y) rectangle +(1,1);
}}
\foreach \x in {3}{
	\foreach \y in {-2}{
		\fill[yslant=0.5, gray] (\x,\y) rectangle +(1,1);
}}
\draw[yslant=0.5,xslant=-1,] (6,3) rectangle +(-3,-3);
\draw[yslant=0.5,xslant=-1] (3,0) grid (6,3);
\foreach \x in {5}{
	\foreach \y in {0}{
		\fill[yslant=0.5,,xslant=-1, gray] (\x,\y) rectangle +(1,1);
}}
\foreach \x in {3}{
	\foreach \y in {2}{
		\fill[yslant=0.5,,xslant=-1, gray] (\x,\y) rectangle +(1,1);
}}
\end{tikzpicture}}
	\subcaptionbox{ \label{Fig:b}}
{\begin{tikzpicture}[on grid, scale=.7]
	\draw[yslant=-0.5,]  (0,2) rectangle +(1,1);
	
	\node  at (1 -0.6, 1.5 -0.2) {$ s_{1} $};
	
	\foreach \x in {0}{
		\foreach \y in {2}{
			\fill[yslant=-0.5, gray] (\x,\y) rectangle +(1,1);
	}}
	
	\draw[yslant=0.5] (1,1) rectangle +(3,1);
	\draw[yslant=0.5] (1,1) grid (4,2);
	
	\foreach \x in {1}{
		\foreach \y in {1}{
			\fill[yslant=0.5, gray] (\x,\y) rectangle +(1,1);
	}}
	
	\draw[yslant=0.5,xslant=-1,] (3,2) rectangle +(3,1);
	\draw[yslant=0.5,xslant=-1] (3,2) grid (6,3);
	
	\foreach \x in {3}{
		\foreach \y in {2}{
			\fill[yslant=0.5,,xslant=-1, gray] (\x,\y) rectangle +(1,1);
	}}
	\end{tikzpicture}}
	\subcaptionbox{ \label{Fig:c}}	
{\begin{tikzpicture}[on grid, scale=.7]
	\draw[yslant=-0.5,]  (0,0) rectangle +(1,3);
	\draw[yslant=-0.5] (0,0) grid (1,1);
	
	\node  at (1 -0.6, -1/2 -0.2) {$ s_{1} $};
	
	\foreach \x in {0}{
		\foreach \y in {2}{
			\fill[yslant=-0.5, gray] (\x,\y) rectangle +(1,1);
	}}
	
	\draw[yslant=0.5] (1,-1) rectangle +(3,3);
	\draw[yslant=0.5] (1,-1) grid (4,2);
	
	\foreach \x in {1}{
		\foreach \y in {1}{
			\fill[yslant=0.5, gray] (\x,\y) rectangle +(1,1);
	}}
	
	\draw[yslant=0.5,xslant=-1,] (3,2) rectangle +(3,1);
	\draw[yslant=0.5,xslant=-1] (3,2) grid (6,3);
	
	\foreach \x in {3}{
		\foreach \y in {2}{
			\fill[yslant=0.5,,xslant=-1, gray] (\x,\y) rectangle +(1,1);
	}}
	\end{tikzpicture}}
\caption{(a) A codeword from an ideal 3-D OOC, black cubes indicate $ 1 $, white indicate $ 0 $. (b) Each of the $ \Lambda S $ space/wavelength sections correspond to a (possibly zero) element from an alphabet of size $ T+1 $. (c) If the code is AMOPP, then each of the $ S $ spatial planes correspond to a (possibly zero) element from an alphabet of size $ \Lambda T +1 $.   }\label{Fig:1}
\end{figure}

Specializing now to ideal codes we observe that a $ (\Lambda\times S\times T,w,0,\la)$ 3D-OOC $ C $  can be viewed as a constant weight ($ w $) code of length $ \Lambda S $ over an alphabet of size $ T+1$ containing zero (See Fig. \ref{Fig:1} (a), (b)). By including the $ T $ distinct cyclic shifts of each codeword we obtain a corresponding constant weight code of size $ T \cdot |C| $. 

It follows that
\begin{equation}\label{eqn: JB}
|C|\le \left \lfloor \frac{A(\Lambda S, w, \la)_{T+1}}{T}\right \rfloor .
\end{equation} 

From Theorem \ref{thm:JB} and the equation (\ref{eqn: JB}) we obtain the following bound for ideal 3-D OOCs.

\begin{thm}\label{thm:ideal3djb}[Johnson Bound for Ideal 3D OOC]\\
	Let $ C $ be an $ (\Lambda\times S \times T,w,0,\la) $-OOC, then 
\begin{align*} 
\Phi(C) & \le J(\Lambda\times S\times  T,w,0,\la_c)\\[1ex]
&= \left\lfloor
\frac{\Lambda S}{w} \left\lfloor \frac{ T(\Lambda S-1)}{w-1}%
\left\lfloor \cdots \left\lfloor \frac{T(\Lambda S-\lambda)}{w-\la}\right\rfloor \right\rfloor \cdots \right\rfloor\right.
\end{align*}      
\end{thm}	

Note that from Theorem \ref{thm:ideal3djb} we see that if $ C $ is an ideal 3D OOC of maximal weight ($w =  \Lambda S $ ) then $ \Phi(C)\le  T^\lambda $

Similarly, (Fig. \ref{Fig:1} (c))  an AMOPP OOC corresponds to a constant weight code of length $ S $ over an alphabet of size $ \Lambda T+1 $ (containing zero). Consequently we obtain the following bound on AMOPP codes. This bound is also found (with a different proof) in \cite{MR3320355}.
 
\begin{thm}\label{thm:AMOPPjb}[Johnson Bound for AMOPP OOC]\\
	Let $ C $ be an $ (\Lambda\times S \times T,w,0,\la) $-AMOPP OOC, then 
	\begin{align*} 
	\Phi(C) & \le \left\lfloor \frac{1}{T}\left\lfloor
	\frac{\Lambda S T}{w} \left\lfloor \frac{ \Lambda T( S-1)}{w-1}%
	\left\lfloor \cdots \left\lfloor \frac{\Lambda T( S-\lambda)}{w-\la}\right\rfloor \right\rfloor \cdots \right\rfloor\right.\right.
	\end{align*} 
\end{thm}	 
 
From the above theorem, we see that if $ C $ is an SPP code (an AMOPP code of maximal weight $ S $) then  $ |C| \le \Lambda^{\la}T^{\la-1} $.\\
Similar reasoning also gives the following two Theorems
\begin{thm}\label{thm:AMOPWjb}[Johnson Bound for AMOPW OOC]\\
	Let $ C $ be an $ (\Lambda\times S \times T,w,0,\la) $-AMOPW OOC, then 
	\begin{align*} 
	\Phi(C) & \le \left\lfloor \frac{1}{T}\left\lfloor
	\frac{\Lambda S T}{w} \left\lfloor \frac{ S T( \Lambda-1)}{w-1}%
	\left\lfloor \cdots \left\lfloor \frac{S T( \Lambda-\lambda)}{w-\la}\right\rfloor \right\rfloor \cdots \right\rfloor\right.\right.
	\end{align*} 
\end{thm}	 
\begin{thm}\label{thm:AMOPTjb}[Johnson Bound for AMOPT OOC]\\
	Let $ C $ be an $ (\Lambda\times S \times T,w,0,\la) $-AMOPT OOC, then 
	\begin{align*} 
	\Phi(C) & \le \left\lfloor \frac{1}{T}\left\lfloor
	\frac{\Lambda S T}{w} \left\lfloor \frac{ \Lambda S( T-1)}{w-1}%
	\left\lfloor \cdots \left\lfloor \frac{\Lambda S( T-\lambda)}{w-\la}\right\rfloor \right\rfloor \cdots \right\rfloor\right.\right.
	\end{align*} 
\end{thm}


Codes meeting the bounds in Theorems \ref{thm: 3djb2} - \ref{thm:AMOPTjb} will be said to be \textit{J-optimal}.
At present, constructions of infinite families of  optimal ideal 3D OOCs are relatively scarce. The codes appearing in the literature seem to be exclusively of the AMOPP or SPP type. According to the bounds established above, it would seem  that for comparable dimensions and weight it may be possible to construct ideal codes with larger capacity than the  AMOPP or AMOPW codes. This is indeed the case. In the following sections we will provide constructions of codes meeting the bounds in Theorem \ref{thm:ideal3djb}.   Table \ref{table:1} will perhaps serve place our constructions in context.

\begin{table}[ht]
	
	\centering
	\caption{ \label{table:1}Summary of known constructions of families of optimal ideal 3D OOC. Unless stated otherwise, $ \la_c=1 $.}
	$p$ a prime, $q $ a prime power, $ \theta(k,q)=\frac{q^{k+1}-1}{q-1} $
	\[
	\begin{array}{|p{8.2cm}|l|c|}
	\hline
	\textrm{ Conditions } & \textrm{ Restrictions }   & \textrm{ Reference}\\
	\hline 
	$ w=S\le p $ for all $ p $ dividing $ \Lambda T $  &    SPP   & \cite{Kim2000} \\
	\hline
	$w= S=\Lambda=T=p $ &  SPP  & \cite{Li2012} \\
	\hline
	$w= S=4\le\Lambda = q $, $ T\ge 2 $&  SPP   &   \cite{Li2012}  \\
	\hline
	$ w= S=q+1,$ $  \Lambda = q>3 $, $ T=p>q $&  SPP   &   \cite{Li2012}  \\
	\hline
	$ w=S=3 $	$ \Lambda,T $ have same parity  & SPP   & \cite{MR3320355}\\
	\hline
	$ w=3 $,  $ \Lambda T(S-1) $ even,  $ \Lambda T(S-1)S\equiv 0 $   mod  $  3$, and \hfill {} \linebreak  $ S\equiv 0,1  $ mod $ 4 $ if $ T\equiv 2 $ mod $ 4 $  and $ \Lambda $ is odd.   &  AMOPP   & \cite{MR3320355}\\
	\hline
	$ w=q+1$,  $  \Lambda S = \theta(m-1,q^{d+1})$,$ T =\theta(d,q)$, $d>0,m>1 $ &     &   \text{Th'm }  \ref{thm:optimal ideal d-spread projective}   \\
	\hline
	$ w=q, $	$  \Lambda S T= q^k-1, T= q-1 $ &     &   \text{Th'm } \ref{thm: codes from affine lines} \\
	\hline
	$ w=q+1, $	$  \Lambda S T= q^2-1, T= q-1, $ $\la_c=q-1 $  &     &   \text{Th'm } \ref{thm:Jopt non-ideal cross} \\
	\hline
	\end{array}
	\]
\end{table}


\section{Preliminaries}\label{sec:Prelim}

Our techniques will rely heavily on the properties of finite projective and affine spaces. Such techniques have been used successfuly in the construction of infinite families of optimal OOCs (for 1-D codes see \cite{MR1022081,MR2359316,MR2067605,MR2354005,AldMellHyperovals}, for 2-D codes see \cite{Alderson20111187,MR2553388}) We start with a brief
overview of the necessary concepts. By $PG(k,q)$ we denote the classical (or Desarguesian) finite projective geometry of dimension $k$ and order $q$. $ PG(k,q) $   may be modeled with the affine (vector) space  $AG(k+1,q)$ of
dimension $k+1$ over the finite field $GF(q)$. Under this model, points of $PG(k,q)$ correspond to 1-dimensional subspaces
of $AG(k,q)$, projective lines correspond to 2-dimensional affine subspaces, and so on.  A \textit{$d$-flat} $\Pi$ in $PG(k,q)$ is a
subspace isomorphic to $PG(d,q)$; if $d=k-1$, the subspace $\Pi$ is called a \textit{hyperplane}.
 Elementary counting  shows that the number of $d$-flats in $PG(k,q)$ is given by the Gaussian coefficient
\begin{equation} \label{eqn: number of projective d-flats} \left[
         \begin{array}{c}
                  k+1 \\
                  d+1 \\
         \end{array}
   \right]_q = \frac{(q^{k+1}-1)(q^{k+1}-q)\cdots(q^{k+1}-q^{d})}{(q^{d+1}-1)(q^{d+1}-q)\cdots(q^{d+1}-q^{d})}
\end{equation}

In particular the number of points of $PG(k,q)$ is given by
$\theta(k,q)=\frac{q^{k+1}-1}{q-1}$.  We will use $\theta(k)$  to represent this number when $ q $ is understood to be the order of the field. Further, we shall denote by $ \LL(k) $ the number of lines in $ PG(k,q) $
.  For a point set $A$ in $PG(k,q)$ we shall denote by $\langle A \rangle$ the span of
$A$, so $\langle A \rangle=PG(t,q)$ for some $t\leq k$.

A \emph{Singer group} of $PG(k,q)$ is a cyclic group of automorphisms acting sharply transitively on the points.  The generator of such a group is known as a \emph{Singer cycle}. Singer groups are known to exist in classical projective spaces of any order and dimension and their existence follows from that of primitive elements in a finite field.

In the sequel we make use of a  Singer group that is most easily understood by modelling a finite projective space using a finite field. If we let $\beta$ be a primitive element
of $GF(q^{k+1})$, the points of $\Sigma=PG(k,q)$ can be represented by the field elements
$\beta^0=1,\beta,\beta^2,\ldots,\beta^{n-1}$ where $n=\theta(k)$. 
The non-zero elements of $GF(q^{k+1})$ form a cyclic group under multiplication.
It is not hard to show that multiplication by $\beta$ induces an automorphism, or
collineation, on the associated projective space $PG(k,q)$ (see e.g. \cite{MR0249317}). Denote by $\phi$ the collineation of
$\Sigma$ defined by $\beta^i \mapsto \beta^{i+1}$. The map $\phi$ clearly acts sharply
transitively on the points of $\Sigma$.

We can construct 3-D codewords by considering orbits under subgroups of $G$.
Let $n=\theta(k)=\Lambda\cdot S \cdot T$ where $G$ is the Singer group of $\Sigma=PG(k,q)$. Since $G$ is cyclic there exists a unique subgroup $H$ of order $T$ ($H$ is the subgroup with generator $\phi^{\Lambda S}$).

\begin{defn}[Projective Incidence Array] \label{defn: Incidence matrices}
Let $\Lambda, S, T$ be positive integers such that $n=\theta(k)=\Lambda\cdot S \cdot T$.  For an arbitrary pointset $\A $ in $\Sigma=PG(k,q)$ we define the $\Lambda\times S \times T$ incidence array  $A=(a_{i,j,k})$, $0\le i\le \Lambda-1$, $ 0\le j \le S-1 $, $ 0\le k \le T-1 $ where $a_{i,j, k}=1$ if and only if the point corresponding to $\beta^{i+j\cdot \Lambda +k\cdot S\Lambda}$ is in $\A$.
\end{defn}

If $\A $ is a pointset of $\Sigma$  with corresponding $\Lambda\times S \times T$ incidence array $A$ of weight $w$, then $\phi^{\Lambda S}$ induces a cyclic
shift on the temporal planes of $A$.  For any such set $\A$, consider its orbit $Orb_H(\A)$ under the group
$H$ generated by $\phi^{\Lambda S}$.  The set ${\A}$ has \textit{full $H$-orbit} if $|Orb_H({\A})|=T=\frac{n}{\Lambda S}$ and \textit{short $H$-orbit} otherwise. If ${\A}$ has full $H$-orbit then a representative member of the orbit and corresponding 3-D codeword is chosen. The collection of all such codewords gives rise to a $(\Lambda\times S \times T,w,\la_a,\la_c)$-3D-OOC, where

\begin{equation}\label{eqn: autocorrelation projective}
\la_a= \max_{0\leq i<j \leq \;T-1} \left\{ |\phi^{\Lambda S\cdot i}({\A})\cap\phi^{\Lambda S\cdot j}({\A})| \right\}
\end{equation}
and
\begin{equation}\label{eqn: crosscorrelation projective}
\la_c=\max_{0\leq i,j \leq \;T-1} \left\{ |\phi^{\Lambda S \cdot i}({\A}) \cap \phi^{\Lambda S\cdot j} ({\A}')| \right\}
\end{equation}
ranging over all ${\A}$, ${\A}'$ with full $H$-orbit.

\subsection{An affine analogue of the Singer automorphism}\label{subsec: affine analogue}
A further automorphism of $\Sigma=PG(k,q)$ shall play a role in our constructions.  It may be viewed as an affine analogue of the Singer automorphism. If a hyperplane $\Pi_{\infty}$ (at infinity) is removed from $PG(k,q)$, what remains is $AG(k,q)$-the $k$-dimensional affine space.  One way to model $AG(k,q)$ is to view the points as the elements of $GF(q^k)$.  Recall that the set $GF(q^k)^*$ of non-zero elements of
$GF(q^k)$ forms a cyclic group under multiplication. Take $\alpha$ to be a primitive element (generator) of $GF(q^k)^*$.  Each nonzero affine point corresponds in the natural way to $\alpha^j$ for some $j$, $0\leq j \leq q^k-2$. Denote by $\psi$ the mapping of $AG(k,q)$ defined by $\psi(\alpha^j)=\alpha^{j+1}$ and $\psi(0)=0$.  The map $\psi$ is an automorphism of $AG(k,q)$ and, moreover, $\psi$ admits a natural extension to an automorphism $\hat{\psi}$ of $PG(k,q)$.  Denote by $\hat{G}$  the group generated by  $\hat{\psi}$.  The fundamental properties of the group $\hat{G}$ central to the constructions here are (for details, see \emph{e.g.} \cite{MR0006735} \cite{MR0249317}.):
\begin{enumerate}
\item  $\hat{G}$  fixes the point $P_0$ corresponding to the field element $0$, and acts sharply transitively on the $q^k-1$ nonzero affine points of $PG(k,q)$.
 \item  $\hat{G}$ acts cyclically transitively on the points of $\Pi_{\infty}$. In particular the subgroup $H=\langle \hat{\psi}^{\theta(k-1)} \rangle$ fixes $\Pi_{\infty}$ pointwise
 .
\end{enumerate}

The 3D-OOCs constructed using affine pointsets will therefore consist of codewords of dimension $\Lambda\times S \times T$, where  $\Lambda\cdot S\cdot T=q^k-1$.

\begin{defn}[Affine Incidence Array] \label{defn: Incidence matrices-affine}
Let $\Lambda, S, T$ be positive integers such that $q^k-1=\Lambda\cdot S\cdot T$.  For an arbitrary pointset $\A$ in $AG(k,q)$ we define the $\Lambda\times S\times  T$ incidence array  $A=(a_{i,j,k})$, $0\le i\le \Lambda-1$, $ 0\le j \le S-1$, $ 0\le k \le T-1$ where $a_{i,j,k}=1$ if and only if the point corresponding to $\alpha^{i+\Lambda  j +S\Lambda k}$ is in $\A$.
\end{defn}

If $\A$ is a set of $w$ nonzero affine points  with corresponding $\Lambda\times S\times  T$ incidence array $A$ of weight $w$, then $\hat{\psi}^{\Lambda S}$ induces a cyclic
shift on the temporal planes of $A$.  For any such set $\A$, consider its orbit $Orb_{\hat{H}}(\A)$ under the group
$\hat{H}=\langle \hat{\psi}^{\Lambda S} \rangle$.  If $\A$ has full $\hat{H}$-orbit then a representative member of the orbit and corresponding 3-dimensional codeword (say $c$) is chosen. The collection of all such codewords  give rise to a $(\Lambda\times S\times T,w,\la_a,\la_c)$-3D-OOC, where

\begin{equation}\label{eqn: autocorrelation affine}
\la_a=\max_{0\leq i<j \leq \;T-1} \left\{ |\hat{\psi}^{S\Lambda\cdot i}(\A)\cap\hat{\psi}^{S\Lambda\cdot j}(\A)| \right\}
\end{equation}
and
\begin{equation}\label{eqn: crosscorrelation affine}
\la_c=\max_{0\leq i,j \leq \;T-1} \left\{ |\hat{\psi}^{S\Lambda \cdot i}(\A) \cap \hat{\psi}^{S\Lambda\cdot j} (\A')| \right\}
\end{equation}

ranging over all $\A$, $\A'$ with full $\hat{H}$-orbit.

\section{Optimal Ideal codes}\label{sec: codes from lines}

\subsection{Codes from projective lines, $ \la_c=1 $}\label{sec: projective lines}

Let $\Sigma=PG(k,q)$ where $G=\langle \phi \rangle $ is the Singer group of $\Sigma$ as in the previous section.  Our work will rely on the following results about orbits of flats.

\begin{thm}[Rao \cite{MR0249317}, Drudge\cite{MR1912797} ]\label{Rao}
In $\Sigma=PG(k,q)$, there exists a short $G$-orbit of $d$-flats if and only if $gcd(k+1,d+1)\ne 1$.
In the case that $ d+1 $ divides $ k+1 $ there is a short orbit $\SSS$ which  partitions the points of $\Sigma$ (i.e. constitutes a $d$-spread of $\Sigma$).  There is precisely one such orbit, and the $G$-stabilizer of any $\Pi\in \SSS$ is $Stab_G(\Pi)=\langle \phi^{\frac{\theta(k)}{\theta(d)}}\rangle$.\\
\end{thm}

Let $\Sigma=PG(k,q)$, $k$ odd with Singer group $G=\langle \phi \rangle$.  Let $\SSS$ be the line spread determined (as in Theorem \ref{Rao}) by $G$  where say $Stab_G(\SSS)=H$.
Consider a line $\ell\notin \SSS$. $\ell$ is incident with precisely $q+1$ members of $\SSS$ and $H$ acts sharply transitively on the points of each line of $\SSS$, so $\ell$ is of full $H$-orbit, that is $|Orb_H(\ell)|=q+1$, and the lines in $Orb_H(\ell)$ are disjoint.   It follows that the number of full $H$-orbits of lines is
\begin{align}\label{eqn: full H-orbits projective}
\text{ \# orbits } & =     \frac{\LL(k) - |\SSS|}{q+1} \nonumber \\ &= \frac{1}{q+1}\cdot \left[\frac{(q^{k+1}-1)(q^{k+1}-q)}{(q^2-1)(q^2-q)}-\frac{\theta(k)}{q+1}\right] \nonumber \\  
   & = \frac{q \cdot\theta(k) \cdot \theta(k-2)}{(q+1)^2} 
\end{align}

For each full $H$-orbit of lines, select a representative member and corresponding (projective)  $ \Lambda \times S \times q+1 $ 3-D incidence array (codeword) where   $\Lambda S = \frac{\theta(k)}{q+1}$ are fixed positive integers.
The collection of all such codewords comprises a $(\Lambda \times S \times (q+1),q+1,\la_a,\la_c)$-3DOOC $C$.  As two lines intersect in at most one point we have (Equation (\ref{eqn: crosscorrelation projective})) $\la_c=1$. Moreover, since the lines in any particular full $H$-orbit $Orb_H(\ell)$ are disjoint, we have (Equation \ref{eqn: autocorrelation projective}) $\la_a=0$.  Hence,   $C$ is a $(\Lambda \times T \times (q+1),q+1,0, 1)$-OOC. From the bound (Theorem \ref{thm:ideal3djb}) we have
\begin{align}\label{eqn: bound projective}
\Phi(C) & = \Phi\left(\Lambda\times S \times (q+1),q+1,0,1\right) \nonumber\\ &
\le  \left\lfloor
\frac{\frac{\theta(k)}{q+1}}{q+1} \left\lfloor \frac{(q+1)(\frac{\theta(k)}{q+1}-1)}{q} \right\rfloor  \right\rfloor = \frac{\theta(k) \cdot q^2 \cdot\theta(k-2)}{q(q+1)^2}
\end{align}
Comparing (\ref{eqn: full H-orbits projective}) and (\ref{eqn: bound projective}) we see that $C$ is in fact optimal.  Noting that $\frac{\theta(k)}{q+1}=\theta(\frac{k-1}{2},q^2)$, we have shown the following.

\begin{thm}\label{thm: codes from proj lines}
Let $q$ be a prime power and let $t\ge 1$.  For any factorisation $ \Lambda S  =  \theta(t,q^2) $ There exists a J-optimal $\left(\Lambda \times S \times (q+1),q+1,0, 1 \right)$-OOC.
\end{thm}

In the codes constructed in Theorem \ref{thm: codes from proj lines}, codewords correspond to lines of $\Sigma = PG(k,q) $ not contained in a particular line-spread. In an analogous way we may generalize whereby codewords correspond to lines that are not contained in any element of a $ d $-spread of $ \Sigma $. We describe this construction as follows.\\
Choose $ d\ge 1 $, $ m>1 $ such that $ k+1=m(d+1) $. Let  $G=\langle \phi \rangle$  be the Singer group as above, and let $\SSS$ be the  $ d $-spread determined (as in Theorem \ref{Rao}) by $G$  where say $Stab_G(\SSS)=H = \left \langle \phi^t \right \rangle$ where $ t=\frac{\theta(k)}{\theta(d)} $.\\
Let $ \Lambda S = t $ be any integral factorization. 
Let  $ \ell $ be a line not contained in any spread element (a $ d $-flat in $ \SSS $), and let $ A $ be the $ \Lambda\times S\times \theta(d)$ projective incidence array corresponding to $ \ell $. As above, $ \ell $ has a full $ H $-orbit. Moreover, as $ H $ acts sharply transitively on the points of each spread element, it follows  that $ A $, when considered as a $ \Lambda\times S\times \theta(d)$ codeword, satisfies $ \la_a = 0 $. For each such line $ \ell $ we choose a representative element of it's $ H $-orbit and include it's corresponding incidence array as a codeword. The aggregate of these codewords gives an ideal $ (\Lambda\times S \times \theta(d), q+1, 0, 1) $-3D OOC,$  C $.  Let us now determine the capacity of $ C $.   Elementary counting shows
\[
\LL(k) = \frac{\theta(k)\theta(k-1)}{q+1}
\] 
We now have 
\begin{align} 
|C| & = \frac{\LL(k)-\LL(d)\cdot \frac{\theta(k)}{\theta(d)}}{\theta(d)} \nonumber  \\  
& = \frac{\theta(k)\theta(k-1)}{\theta(d)(q+1)} - \frac{\theta(d-1)\theta(k)}{\theta(d)(q+1)} \nonumber \\
& = \frac{\theta(k)}{\theta(d)(q+1)}\left[\theta(k-1)-\theta(d-1)\right] \label{eqn: size d-spread proj}
\end{align}
From Theorem \ref{thm:ideal3djb} we have the corresponding Johnson Bound is

\begin{align} 
\Phi(C) & \le \left\lfloor \frac{\frac{\theta(k)}{\theta(d)}}{q+1} \left\lfloor \frac{\theta(d) \left(\frac{\theta(k)}{\theta(d)}-1\right)}{q}\right\rfloor\right\rfloor \nonumber \\  
& = \frac{\theta(k)}{\theta(d)(q+1)}\left(\frac{\theta(k)-\theta(d)}{q}\right) \nonumber \\
& = \frac{\theta(k)}{\theta(d)(q+1)}\left[\theta(k-1)-\theta(d-1)\right] \label{eqn: JB d-spread proj}
\end{align}

Comparing (\ref{eqn: size d-spread proj}) and (\ref{eqn: JB d-spread proj}) we see the codes obtained are J-optimal. With the observation that  $ \frac{\theta{(k)}}{\theta(d)} =\theta(m-1,q^{d+1}) $, we  have shown the following.
\begin{thm}\label{thm:optimal ideal d-spread projective}
For $ d\ge 1 $, $ m>1 $, and $ \Lambda S =\theta(m-1,q^{d+1}) $, there exists a J-optimal $ (\Lambda\times S\times\theta(d), q+1, 0, 1) $-OOC .	 
\end{thm} 

\subsection{Codes from projective planes, $ \la_c=q-1 $}\label{sec: Codes from affine lines}
Let $\Sigma=PG(3,q)$,  with Singer group $G=\langle \phi \rangle$ as in the previous section.  Let $\SSS$ be the line spread determined  by $G$  where say $Stab_G(\SSS)=H = \langle \phi^{\frac{\theta(3)}{q+1}} \rangle $. For each line $ \ell\in \SSS $ select a plane $ \Pi_\ell $ containing $ \ell $. Let $ \Pi^*_\ell =\Pi_\ell \setminus \ell $, so that in particular each $ \Pi^*_\ell $ comprises $ q^2 $ coplanar points. Fix a factorization $ \Lambda S = q^2+1 $ and for each $ \ell \in \SSS $  let $ A_\ell $ be the $\Lambda \times S \times q+1 $ projective incidence array  of $ \Pi^*_\ell $.  We claim that the aggregate of these codewords  constitutes a $ (\Lambda\times S \times q+1, q^2,0, q-1 ) $ 3-D OOC, $ C $. 
The dimensions and weight are clear. Note that since every plane has full $ G $ orbit, each $ \Pi_\ell $ has full $ H $-orbit. Any two planes $ PG(3,q) $ meet (precisely) in a line, and any two lines in a projective plane must meet in a point. As such, $ \la_a=0 $ follows from the fact that for any $ \ell \in \SSS $ and for any non-identity $ \gamma\in H $
\[
\Pi_\ell \cap \gamma(\Pi_\ell) =\ell
\]
and therefore
\[
\Pi^*_\ell \cap \gamma(\Pi^*_\ell) = \emptyset
\]
giving $ \la_a=0$.
For the auto-correlation, suppose  $ \Pi^*_\ell $ and $ \Pi^*_{\ell'} $ correspond to (any cyclic shift of) two codewords, so that $ m = \Pi_\ell \cap \Pi_{\ell'} $ is a line containing a point   $ P $ of $ \ell $ and a point $ P' $ of $ \ell' $. As $ \ell $  and $ \ell' $ are skew we have $ P\ne P' $,  so we have
\begin{equation}
\left|\Pi^*_{\ell} \cap \Pi^*_{\ell'}\right| =q-1,
\end{equation}  
giving $ \la_c=q-1$. Finally, since   $ |C|= | \SSS|= q^2+1=\Lambda S$ and $ w^2=q^4>q^4-1=\Lambda S T \la_c $, the bound (\ref{eqn:3djbb}) in Theorem \ref{thm: 3djb2} shows $ C $ to be optimal. We have shown the following.
\begin{thm}\label{thm:Jopt non-ideal cross}
If $ q $ is  a prime power and $ \Lambda S=q^2+1$, then there exists a J-optimal $ (\Lambda \times S \times q+1, q^2,0,q-1) $-3D OOC. 	
\end{thm} 
Though this construction does not generally produce codes of small cross-correlation, it does produce  an infinite family of optimal codes with ideal auto-correllation meeting the second bound (\ref{eqn:3djbb}) in Theorem \ref{thm: 3djb2}. Thus, the bound (\ref{eqn:3djbb}) is sometimes tight with $ \la_c>1 $.

\subsection{Ideal Codes from Affine Lines, $ \la_c=1 $}\label{sec: Codes from affine lines}

Let $\Sigma=PG(k,q)$ where $E=\Sigma\setminus \Pi_{\infty}$ is the associated affine space $AG(k,q)$. Let $\hat{G}=\langle \hat{\psi} \rangle $ be the map as
described in Section \ref{subsec: affine analogue} based on the primitive element $\alpha$ of $GF(q^k)^*$.  Our affine analog of Theorem \ref{Rao} follows from Theorem 8 of \cite{MR0249317}.

\begin{thm}[Rao \cite{MR0249317}]\label{Rao affine d-flats}
A $d$-flat $\Pi$ in $PG(k,q)$ is of full $\hat{G}$-orbit if and only if  the origin $P_0\notin \Pi$ and $\Pi$ is not a subset of $\Pi_{\infty}$.
\end{thm}

From the Theorem  \ref{Rao affine d-flats} it follows that each point of $\Pi_\infty$ is incident with precisely $q^{k-1}-1$ lines of full $\hat{G}$-orbit. Let $\hat{H}=\langle \hat{\psi}^{\theta(k-1)} \rangle$ be the unique subgroup of order $q-1$. Note that $\hat{H}$ fixes each point of $\Pi_\infty$.
Clearly, any line with full $\hat{G}$-orbit is also of full $\hat{H}$-orbit.  The number of full $\hat{H}$-orbits of  lines is therefore at least
\begin{equation}\label{eqn: size of affine line code}
\frac{\theta(k-1)\cdot (q^{k-1}-1)}{q-1}=\theta(k-1)\cdot\theta(k-2).
\end{equation}

Let $ \Lambda S = \theta(k-1) $ be any fixed factorisation. For each full $\hat{H}$-orbit, select a representative line $\ell$ and corresponding (affine) $\Lambda\times S \times \times (q-1)$ incidence array $A$ (corresponding to the points of $\ell'=\ell\cap E$), a $(\Lambda\times S \times (q-1), w, \la_a,\la_c)$-3D-OOC $C$ results.

Each representative line $\ell$ used in the construction meets $\Pi_\infty$ in precisely one point, say $\ell\cap\Pi_\infty=P_\infty$, so codewords are of weight $q$.
As two lines meet in at most one point we get $\la_c=1$.  Moreover, since $P_\infty$ is fixed under the action of $\hat{H}$, the orbit $Orb_{\hat{H}}(\ell)$ comprises $|H|=q-1$  lines, each incident with $P_\infty$ (in particular, no two meet in an affine point). Therefore, we have $\la_a=0$ and  $|C|$ is given by (\ref{eqn: size of affine line code}).\\
From Theorem \ref{thm:ideal3djb} we have
\begin{align}
\Phi(C) & = \Phi(\Lambda\times S \times (q-1),q,1)  \nonumber \\ 
& \le \left\lfloor \frac{\theta(k-1)}{q} \left\lfloor \frac{q^k-2}{q-1} \right\rfloor  \right\rfloor  \nonumber \\
& = \left\lfloor \theta(k-1)\cdot\theta(k-2) \right\rfloor=|C|
\end{align}

We have shown the following

\begin{thm} \label{thm: codes from affine lines}
For $q$ a prime power, for each $t$, and for any factorisation $ \Lambda S = 
\theta(t) $ there exists a J-optimal $(\Lambda \times S\times (q-1),q,0,1)$-OOC.
\end{thm}

\section{Conclusion}
In this paper we provided several constructions of infinite families of 3-dimensional OOC's. In each case the families have ideal autocorrelation $ \la_a=0 $ and are optimal with respect to the Johnson bounds presented or developed here. A key feature of the constructions presented  involve two or more parameters that may grow without bound and at each stage produce optimal codes.


%



\section*{Acknowledgment}

The author acknowledges support from the NSERC of Canada.

\ifCLASSOPTIONcaptionsoff
  \newpage
\fi

\begin{IEEEbiographynophoto}{Tim Alderson}
Biography text here.
\end{IEEEbiographynophoto}






\end{document}